\documentclass[apl,superscriptaddress,reprint] {revtex4-1}
\usepackage{graphicx}
\usepackage{amsmath}
\usepackage{natbib}

\begin{document}

\title{Intrinsic Origin of Negative Fixed Charge in Wet Oxidation for Silicon Carbide}
\author{Yasuhiro Ebihara} 
\affiliation{Graduate School of Pure and Applied Sciences, University of Tsukuba, 1-1-1 Tennodai, Tsukuba, Ibaraki, 305-8571, Japan}
\author{Kenta Chokawa} 
\affiliation{Graduate School of Pure and Applied Sciences, University of Tsukuba, 1-1-1 Tennodai, Tsukuba, Ibaraki, 305-8571, Japan}
\author{Shigenori Kato} 
\affiliation{Graduate School of Pure and Applied Sciences, University of Tsukuba, 1-1-1 Tennodai, Tsukuba, Ibaraki, 305-8571, Japan}
\author{Katsumasa Kamiya} 
\email{kkamiya@comas.frsc.tsukuba.ac.jp}
\affiliation{Graduate School of Pure and Applied Sciences, University of Tsukuba, 1-1-1 Tennodai, Tsukuba, Ibaraki, 305-8571, Japan}
\author{Kenji Shiraishi}
\affiliation{Graduate School of Pure and Applied Sciences, University of Tsukuba, 1-1-1 Tennodai, Tsukuba, Ibaraki, 305-8571, Japan}
\date{\today}

\begin{abstract}
We demonstrate on the basis of first-principles calculations that the formation of carbonate-like moiety in SiO$_2$ could be the intrinsic origin of negative fixed charge in SiC thermal oxidation. We find that two possible origins for the negative fixed charges are O-lone-pair state and a negatively charged CO$_3$ ion in SiO$_2$. Such CO$_3$ ion is able to be formed as a result of the existence of residual C atoms in SiO$_2$, which are expected to be emitted from the interface between SiC and SiO$_2$, and the incorporation of H atoms during wet oxidation. 
\end{abstract}
\maketitle

Silicon carbide (SiC) has recently attracted a great deal of attention as a promising material for the next generation of high-power and high-frequency electronic devices. One of advantages of SiC for practical application is the formation of insulating silicon oxide (SiO$_2$) layers on SiC by thermal oxidation, being important for the fabrication of SiC-based metal-oxide-semiconductor field-effect transistors (MOSFETs) \cite{pantelides_2006,dimitrijev 2003,afanasev 1997}. There are two types of oxidation processes; dry and wet oxidation. The dry oxidation process of SiC is performed in O$_2$ atmosphere. On the other hand, the wet oxidation process is conducted in O$_2$ and H$_2$O atmosphere, and, as compared to the case of dry oxidation, the wet oxidation process is known to effectively reduce the deep donor-type interface states for the $p$-type MOS capacitors, leading to the improvement of the channel mobility of the MOSFET \cite{yano 1999,okamoto 2006}. Furthermore, wet oxidation yields high oxidation speed, twenty five times larger than that of dry oxidation, being also advantageous for realistic applications toward mass productions \cite{benfdila 2010}.

It is, however, serious issue encountered in wet oxidation. This treatment leads to surprisingly large amount of negative fixed charges near the SiC/SiO$_2$ interfaces \cite{yano 1999,palmieri 2009}. The wet-oxidation-induced interface traps between SiC and thermal oxide SiO$_2$ have been investigated experimentally by high-frequency C-V measurement of MOS capacitors \cite{inoue 1997,yano 1999}. It has been reported that fixed charges near SiO$_2$/SiC interfaces give large influences on threshold voltage and cannel mobility of SiC-MOSFET. Notably, it has been pointed out that in wet oxidation the interface fixed charges increase up to $-$13$\times$10$^{11}$ cm$^{-2}$ (wet reoxidation annealed after wet oxidation) in $p$-4H-SiC MOS capacitor \cite{yano 1999}. Non-negligible negative fixed charge also appears during dry oxidation \cite{watanabe 2011,fukuda 2000} . Nevertheless, despite the extensive experimental outcomes, the origin of the negative fixed charge remains unclear for the last decade.

In this letter, we propose to clarify the atomistic origin of negative fixed charge appearing in SiC thermal oxidation processes on the basis of first-principles calculations in the framework of density functional theory (DFT). We here focus on residual C atoms in SiO$_2$ which are expected to be emitted from the interface between the SiC and SiO$_2$. The existence of such residual C atoms is supported through a comparison with Si thermal oxidation. Indeed, it has been reported experimentally and theoretically that one-third Silicon atoms are inevitably emitted from the interface to release the stress induced during Silicon oxide growth \cite{kageshima 1998,ming 2006}. 

Our calculations demonstrate that residual C atoms in SiO$_2$ leads to the formation of carbonate-like ion assisted by incorporated H atoms. We find that the existence of the carbonate-like ion stabilizes 1$-$ charge state, being the intrinsic origin of negative fixed charge in SiO$_2$/SiC interface.
  
We considered three types of structural models based on 72-atoms $\alpha$-quarts SiO$_2$. The first model was constructed by substituting one C atom for Si, while the second and third models involve one and three H atoms, respectively. The lattice vectors were set to be $a_1 = (1, -\sqrt{3}, 0)a$, $a_2 = (1, \sqrt{3}, 0) a$, and $a_3 = (0, 0, 2) c$, where $a=4.944$ \r{A} and $c=5.363$ \r{A}, which was optimized for SiO$_2$ bulk in the $\alpha$-quarts phase. All calculations were performed using DFT with the Perdew-Burke-Ernzerhof generalized gradient approximation \cite{kohn 1965,yamauchi 1996,perdew 1996}. We used ultrasoft pseudopotentials for Si, O and C and normconserving pseudopotentials for H \cite{troullier 1991,vanderbilt 1990}. We used a plane-wave basis set with a cutoff energy of 36 Ry. We sampled 2$\times$2$\times$2 $k$-points in the Brillouin zone (BZ) for the BZ integration; using 3$\times$3$\times$3 $k$-points in the BZ gives a difference of 0.01 eV in total-energy for the system, assuring required accuracy in the present calculations. After the optimizations, all the atomic forces were less than 0.05 eV/\r{A}. We used the background ion charge correction proposed by Bl\"ochl to calculate the charged state \cite{blochl 2000}. 

\begin{figure*}[tbl]
\begin{center}
\includegraphics[width=160mm]{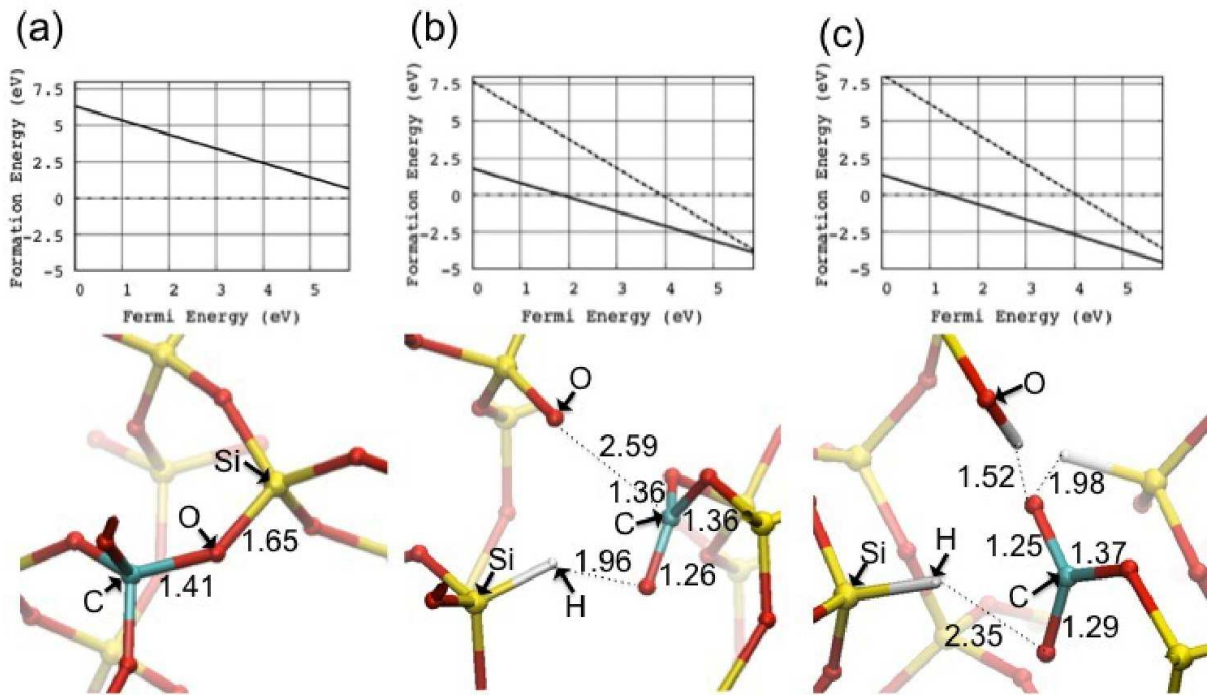}
\end{center}
\caption{
(Color online)
Formation energy diagram and optimized structures for the models including (a) one C atom, (b) one C atom and one H atom, and (c) one C atom and three H atoms in the unit cell. Dashed, solid, and dotted lines in the formation energy diagram show neutral, $1-$, and $2-$ charge state, respectively. Optimized structures are shown for neutral state in the panel (a), and for $1-$ charge state for the panels (b) and (c). Distances are measured in units of \AA.
}
\label{eform}
\end{figure*}

Figure \ref{eform} shows the the formation energy diagram of the models including one C atom, one C atom and one H atom, and one C atom and three H atoms, respectively. We found that a defect with only C atom takes the neutral charge state for the whole range of the Fermi energy (the upper panel in Fig. \ref{eform} (a)). However, the system including H atoms takes the 1$-$ charge state for nearly the whole range of the Fermi energy (the upper panels in Fig. \ref{eform} (b) and (c)). Particularly, if we assume that the valence band offset parameter between 4H-SiC and SiO$_2$ is 2.9 eV \cite{afanasev 2000}, the 1$-$ charge state appears as the most stable state for all over the range of SiC band gap region. These results indicate that negative fixed charge is formed when both of C and H atoms exist.

Notably, we found that the structural origin of the stabilization for 1$-$ charge state is the formation of carbonate-like ion. The lower panels in Fig. \ref{eform} shows the optimized structures for each model. It is clearly shown in the figure that carbonate-like (CO$_3$-like) moiety is formed in SiO$_2$ in the H-atom-included cases (Fig. \ref{eform} (b) and (c)). The formation of CO$_3$-like configuration is ascribed to H atoms that can cut Si$-$O bonds near the substitutional C atom, leading to the formation of strong sp$^2$-like C-O bonds. These results provide clear indication that C atoms originating from SiO$_2$/SiC interface leads to the formation of sp$^2$ network in SiO$_2$, which is assisted by H atoms.

There are two main factors for the stabilization of the CO$_3$-like moiety. The first one is partial resonance in the CO$_3$-like configuration. Indeed, as shown in the lower panels of Fig. \ref{eform} (b) and (c), its C-O bonds are characterized as partially double bond and fully double bond, as compared with the typical lengths of the C-O single and double bonds (1.42 and 1.21 \r{A}, respectively). The second factor is the formation of hydrogen-bonding between the CO$_3$-like part and the nearby Si-O-H moiety (Fig. \ref{eform} (c)).

\begin{figure}[tbl]
\begin{center}
\includegraphics[width=80mm]{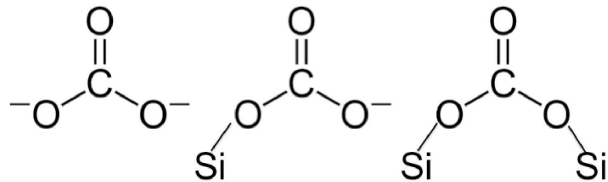}
\end{center}
\caption{
Schematic illustration of structural change for CO$_3$-like moiety.
}
\label{co3}
\end{figure}
The detailed structure of the CO$_3$-like moiety depends on the number of incorporated H atoms in the system. In the case of the one H incorporation, two of three O atoms in the CO$_3$-like part form covalent bonds with Si atoms (the lower panel in Fig. \ref{eform} (b)). The CO$_3$-like structure has C-O bonds of 1.26, 1.36, and 1.36 \r{A}, and O-C-O angles of 113, 122, and 125 degrees. On the other hand, in the case of the incorporation of three H atoms, only one O atom in the CO$_3$-like moiety form a covalent bond with Si atom (the lower panel in Fig. \ref{eform} (c)). In this case, C-O bonds become 1.27, 1.29, and 1.36 \r{A}, and O-C-O angles of 116, 121, and 123 degrees. These structural change of CO$_3$-like moiety can be understood by considerations about its local bond network (Fig. \ref{co3}). As compared with an isolated CO$_3$-like ion, Si-O bond formation leads to a structural change of CO$_3$-like moiety, depending on the number of Si-O bonds. This chemical picture also strongly suggests that CO$_3$-like part is in $1-$ charge state when one Si-O bond is formed, while it is in neutral charge state when two Si-O bonds are created.

\begin{figure}[tbl]
\begin{center}
\includegraphics[width=85mm]{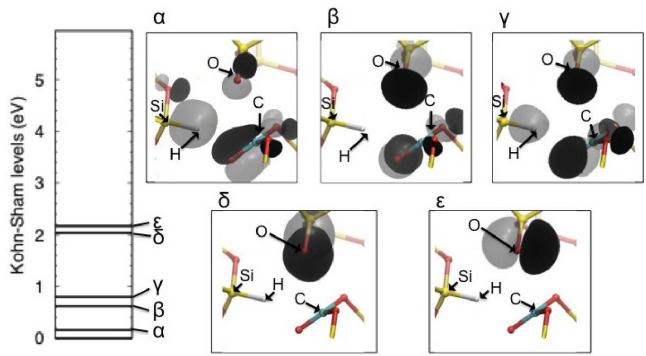}
\end{center}
\caption{
(Color online) Energy levels around energy gap at the $\Gamma$ point and the distribution of wave functions for the system including one C atom and one H atom in $1-$ charge state. Energies are measured from the SiO$_2$ valence band top. The isovalues for the wave-function distribution are $+$ 0.12 (gray) and $-$ 0.12 (black) ($e$/\AA$^3$)$^{1/2}$.
}
\label{wave1}
\end{figure}
Indeed, a detail analysis of the electronic structure reveals that there are two origins for negative fixed charge, depending on the number of incorporated H atoms in the system. Figure \ref{wave1} shows the energy level and the distribution of wave functions for the system including only one H atom in $1-$ charge state. There are five occupied gap states named $\alpha$, $\beta$, $\gamma$, $\delta$, and $\epsilon$ in the SiO$_2$ band gap region. The energy levels for the $\alpha$, $\beta$ and $\gamma$ states appear in the range of $\sim$ 1 eV from the SiO$_2$ valence band top. These states are distributed primarily on CO$_3$-like moiety and the nearby O atom. On the other hands, the $\delta$ and $\epsilon$ states emerge above $\sim$ 2 eV from the SiO$_2$ valence band top. These states are distributed only on O atom with lone-pair characteristics. These results indicate clearly that CO$_3$-like moiety is in neutral charge state, and an extra electron is trapped into O-atom lone-pair electron state, which leads to the intrinsic origin of negative fixed charge. 

\begin{figure}[tbl]
\begin{center}
\includegraphics[width=85mm]{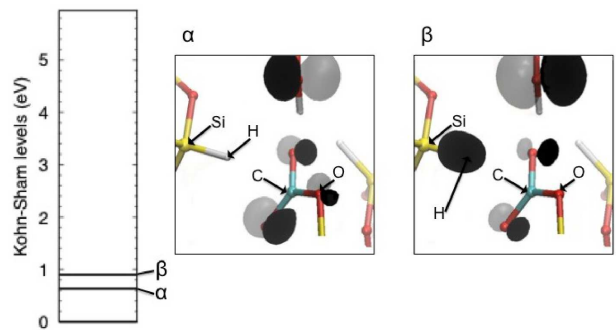}
\end{center}
\caption{
(Color online) Energy levels around energy gap at the $\Gamma$ point and the distribution of wave functions for the system including one C atom and three H atoms in $1-$ charge state. Energies are measured from the SiO$_2$ valence band top. The isovalues for the wave-function distribution are $+$ 0.12 (gray) and $-$ 0.12 (black) ($e$/\AA$^3$)$^{1/2}$.
}
\label{wave2}
\end{figure}
In sharp contrast, the three-H-atom incorporation case exhibits completely different electronic structure. There are only two gap states appearing above $\sim$ 1 eV from the SiO$_2$ valence band top. The corresponding wave functions are distributed strongly in CO$_3$-like moiety. This indicates clearly that an extra electron is trapped into the CO$_3$-like part, and the moiety is in $1-$ charge state. The CO$_3$-like ion is thus the intrinsic origin of negative fixed charge when enough H atoms are incorporated in the system, and this could be possible in wet oxidation. In particular, the energy level for this defect state is deep as compared with SiO$_2$ large gap. This stabilization is ascribed to the formation of hydrogen-bonding between the CO$_3$-like ion and the nearby Si-O-H moiety as mentioned above. These results strongly suggest that the defect state originated from CO$_3$-like ion does not become a charge trapping level, but plays a strong negative-fixed-charge appearing in wet oxidation. 

On the other hand, it has been experimentally reported that a few but non-negligible negative fixed charges appear during dry oxidation \cite{watanabe 2011,fukuda 2000}. In this case, the negative fixed charge could be O-lone-pair state. In reality, SiO$_2$ structure is amorphous. It is thus expected that a few O atoms which is not terminated by H can be formed near incorporated C atoms. This situation is similar to the case shown in Fig. \ref{eform} (b), where lone-pair electron state originated from non-terminated O-atom appears in SiO$_2$ band gap (Fig. \ref{wave1}).

\begin{figure}[tbl]
\begin{center}
\includegraphics[width=65mm]{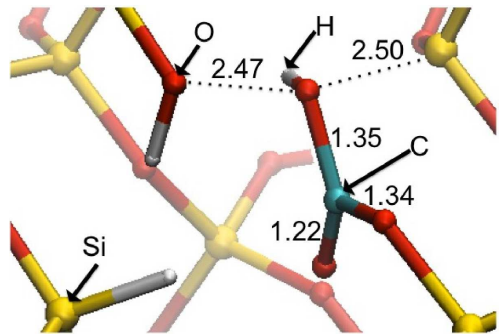}
\end{center}
\caption{
(Color online) Optimized structure for the system including one C atom and three H atoms in $1-$ charge state. Distances are measured in units of \AA.
}
\label{neutral}
\end{figure}
Finally, we note charge-state-dependent behavior of H atoms near CO$_3$-like ion. As shown in Fig. \ref{eform} (c), a H atom terminates Si dangling bond when CO$_3$-like ion is formed. This H atom is, however, transferred from Si atom to one O atom in the CO$_3$-like ion when the system is in neutral charge state (Fig. \ref{neutral}), leading to the formation of Si dangling bond.

In conclusion, we have demonstrated that the formation of carbonate-like moiety in SiO$_2$ dielectrics could be the origin of negative fixed charges in the SiC thermal oxidation using first-principles calculations in the framework of DFT. Our calculations showed two possible atomistic origins for the negative fixed charges: O-lone-pair state and a negatively charged CO$_3$ ion. The CO$_3$ ion is able to be formed as a result of the existence of residual C atoms in SiO$_2$, which are expected to be emitted from the interface between SiC and SiO$_2$, and the incorporation of H atoms during wet oxidation. 

Computations were performed on a NEC SX-9 at the Institute for Solid-State Physics, The University of Tokyo, and a Fujitsu PrimeQuest at the Research Center for Computational Science, Okazaki Research Facilities, National Institutes of Natural Sciences. This research was supported by a Grant-in-Aid for Young Scientists (B) (No. 22740259) from the Japan Society for the Promotion of Science.


\end{document}